\documentclass{elsart}
\usepackage{times}
\usepackage{epsfig}
\usepackage{marvosym}
\usepackage{amsmath}
\usepackage{amsfonts}
\usepackage{amssymb}
\begin{document}
\bibliographystyle{roman}

\journal{Nuclear Instruments and Methods A}

\def\hb{\hfill\break}
\def\MeV{\rm MeV}
\def\GeV{\rm GeV}
\def\TeV{\rm TeV}

\def\m{\rm m}
\def\cm{\rm cm}
\def\mm{\rm mm}
\def\lam{$\lambda_{\rm int}$}
\def\rad{$X_0$}
 
\def\NIM{Nucl. Instr. and Meth.~}
\def\ieee {{IEEE Trans. Nucl. Sci.~}}

\def\etal{{\it et al.}}
\def\eg{{\it e.g.,~}}
\def\ie{{\it i.e.,~}}
\def\cf{{\it cf.~}}
\def\etc{{\it etc.~}}
\def\vs{{\it vs.~}}

\hyphenation{ca-lo-ri-me-ter}
\hyphenation{ca-lo-ri-me-ters}

\begin{frontmatter}
\title{Measurement of the Contribution of Neutrons to Hadron Calorimeter Signals}

\author{N. Akchurin$^a$, L. Berntzon$^a$, A. Cardini$^b$, R. Ferrari$^c$, G. Gaudio$^c$,}
\author{J. Hauptman$^d$, H. Kim$^a$, L. La Rotonda$^e$, M. Livan$^c$, E. Meoni$^e$,}
\author{H. Paar$^f$, A. Penzo$^g$, D. Pinci$^h$, A. Policicchio$^e$,} 
\author{S. Popescu$^{i,}$\thanksref{Leave}}
\author{G.~Susinno$^e$, Y. Roh$^a$, W. Vandelli$^c$ and R. Wigmans$^{a,}$\thanksref{Corres}}

\address{$^a$ Texas Tech University, Lubbock (TX), USA\\
$^b$ Dipartimento di Fisica, Universit\`a di Cagliari and INFN Sezione di Cagliari, Italy\\
$^c$ Dipartimento di Fisica Nucleare e Teorica, Universit\`a di Pavia and INFN Sezione di Pavia, Italy\\
$^d$ Iowa State University, Ames (IA), USA\\
$^e$ Dipartimento di Fisica, Universit\'a della Calabria and INFN Cosenza, Italy\\
$^f$ University of California at San Diego, La Jolla (CA), USA\\
$^g$ INFN Trieste, Italy\\
$^h$ Dipartimento di Fisica, Universit\`a di Roma ''La Sapienza''  and INFN Sezione di Roma\\
$^i$ CERN, Gen\`eve, Switzerland}
\thanks[Leave]{On leave from IFIN-HH, Bucharest, Romania.}
\thanks[Corres]{Corresponding author.
              Email wigmans@ttu.edu, fax (+1) 806 742-1182.}

\begin{abstract}
The contributions of neutrons to hadronic signals from the DREAM calorimeter are measured by
analyzing the time structure of these signals. The neutrons, which mainly originate from the evaporation stage of nuclear breakup in the hadronic shower development process, contribute through
elastic scattering off protons in the plastic scintillating fibers which provide the $dE/dx$ information in this calorimeter. This contribution is characterized by an exponential tail in the pulse shape, with a time
constant of $\sim 25$ ns. The relative contribution of neutrons to the signals increases with the distance from the shower axis.
As expected, the neutrons do not contribute to the DREAM \v{C}erenkov signals.
\end{abstract}

\begin{keyword}
Calorimetry, \v{C}erenkov light, optical fibers, neutron signals
\end{keyword}
\end{frontmatter}

\section{Introduction}

The energy resolution of calorimeters is determined by fluctuations. To improve  that
resolution significantly, one has to address the dominating source of these fluctuations. In non-compensating calorimeters, fluctuations in the electromagnetic shower fraction  ($f_{\rm em}$) dominate the energy resolution for hadrons and jets. These fluctuations, and their energy-dependent characteristics, are also responsible for other undesirable calorimeter characteristics, in particular hadronic signal non-linearity and a non-Gaussian response function.

The DREAM (Dual-REAdout Method) calorimeter was developed as a device that would make it possible to eliminate the effects of these fluctuations. The detector is based on a copper absorber structure, equipped with two types of active media which measure 
complementary characteristics of the shower development. Scintillating fibers measure the total energy deposited
by the shower particles, while \v{C}erenkov light is only produced by the charged, relativistic shower particles.
Since the latter are almost exclusively found in the electromagnetic (em) shower component (dominated by
$\pi^0$s produced in hadronic showers), a comparison of the two signals makes it possible to measure $f_{\rm em}$, event by event. As a result, the effects of fluctuations
in this component can be eliminated. This leads to an important improvement in the
hadronic calorimeter performance.
The performance characteristics of this detector are described elsewhere \cite{DREAMem,DREAMhad,DREAMmu}. 

The elimination of (the effects of) the dominant source of fluctuations means that other types of fluctuations now dominate the detector performance. Further improvements may be obtained by concentrating on these. The energy resolution of a  calorimeter of the type described above is limited by
fluctuations in the \v{C}erenkov light yield  and by sampling fluctuations. In another paper, we demonstrate
that these effects may be effectively reduced by using a homogeneous calorimeter that produces a (separable) mixture of scintillation and \v{C}erenkov light \cite{DREAM3cal}.

Once the mentioned effects have been eliminated, the performance of this type of detector may approach the theoretical hadronic energy resolution limit. This limit is determined by 
so-called fluctuations in {\sl visible energy}, which result from the fact that some (variable) fraction of the energy carried by the showering particle is used to provide the nuclear binding energy needed to release nucleons and nucleon aggregates in nuclear reactions. This energy does itself not result in a measurable signal. However, it has been shown that efficient detection of the neutrons abundantly produced in these processes may be an effective tool for reducing the (effects of) fluctuations in visible energy, and that hadronic energy resolutions of $15-20\%/\sqrt{E}$ might be achieved this way \cite{RWbook}. In the present paper, we report on efforts to measure the contribution of neutrons to the signals of the DREAM calorimeter.

In Section 2, we briefly discuss the reasons for the importance of neutrons in this context, and we describe the experimental technique we used to get a handle on their contributions to the calorimeter signals.
In Section 3, we describe the calorimeter and the
experimental setup in which it was tested.  Experimental
results are presented and discussed in Section 4.  Summarizing conclusions are given in Section 5.  

\section{Neutrons in hadron calorimetry} 

\subsection{The role of neutrons} 

When an incoming high-energy hadron strikes an atomic nucleus, the most likely process to
occur is spallation. Spallation is usually described as a two-stage process: a fast
intranuclear cascade, followed by a slower evaporation stage. The incoming hadron makes
quasi-free collisions with nucleons inside the struck nucleus. The affected nucleons
start traveling themselves through the nucleus and collide with other nucleons. In
this way, a cascade of fast nucleons develops. 
In this stage, also pions and other
unstable hadrons may be created if the transferred energy is sufficiently high. Some of
the particles taking part in this cascade reach the nuclear boundary and escape. Others
get absorbed and distribute their kinetic energy among the remaining nucleons in the
nucleus, resulting in the production of an excited intermediate nucleus.

The second step of the spallation reaction consists of the de-excitation of this
intermediate nucleus. This is achieved by evaporating a certain number of particles,
predominantly free nucleons, but sometimes also $\alpha$'s or even heavier nucleon
aggregates, until the excitation energy is less than the binding energy of a single nucleon.
The remaining energy, typically a few MeV, is released in the form of $\gamma$ rays.
In very heavy nuclei, \eg\ uranium, the intermediate nucleus may also fission.

In these spallation reactions, considerable numbers of nucleons may be released from the
nuclei in which they are bound.
The energy needed to release these nucleons, \ie\ the nuclear binding energy, is lost for
calorimetric purposes. It does not contribute to the calorimeter signal, and is thus called
{\em ``invisible''}. 

There is a large variety of processes that may occur in hadronic shower
development and event-to-event fluctuations in the invisible energy fraction are substantial.
On average, invisible energy accounts for 30-40\% of the non-em
shower energy, \ie\ energy that is not carried by
$\pi^0$s or other electromagnetically decaying particles produced in the shower
development \cite{RSI}.

The large event-to-event fluctuations in visible energy have obviously direct
consequences for the precision with which hadronic energy can be measured in
calorimeters. Because of these fluctuations, which have no equivalent in electromagnetic
shower development processes, the energy resolution with which hadron showers can be measured
is usually considerably worse than the resolution with which em showers can be measured. 
There is, however, an elegant way in
which one can limit these effects, by exploiting the correlation that exists between the
invisible energy lost in releasing nucleons from the nuclei in which they are bound and
the kinetic energy carried by these nucleons \cite{Acosta}.

As indicated above, the nucleons produced in spallation reactions can be divided into
two classes, the {\em cascade} and the {\em evaporation} nucleons. The energy spectrum of
the latter is considerably softer and most of these nucleons are neutrons. This is because
the Coulomb barrier prevents soft protons from being released and also because of the
larger abundance of neutrons in target nuclei. The evaporation neutrons  carry an average
%
%
kinetic energy of 2 - 3 MeV 
The cascade nucleons are more energetic, but also much less numerous than the
evaporation nucleons. Therefore, the vast majority of the nucleons released in hadronic shower development
are {\sl neutrons}, especially in high-$Z$ materials.

Experimental measurements 
have revealed that the numbers of neutrons produced in hadronic shower development are
large, \eg 20 per GeV in lead absorber \cite{Leroy}.
There is a clear correlation between the
total amount of invisible energy (\ie\ the number of target nucleons released in the
development of the hadron shower) and the total kinetic energy carried by these neutrons. This correlation can be exploited to achieve a substantial improvement of the calorimetric
performance \cite{RSI,Acosta}. 

\subsection{How to detect shower neutrons?}

Shower neutrons can only be detected through the effects of the nuclear reactions they initiate.
The most abundant nuclear evaporation neutrons are produced with typical kinetic energies of a few MeV.
At these energies, the most important process through which they lose this kinetic energy is {\sl elastic scattering}.
After they have lost practically all their kinetic energy, the neutrons may be captured by a nucleus, and generate
typically 7 - 8 MeV in the form of nuclear  $\gamma$s, when this compound nucleus falls back to its ground state. 
However, since the thermalization process that preceeds this capture takes typically at least $1 \mu$s, the latter process is in practice not interesting for calorimetric purposes \cite{zeus}.

Since the average energy fraction transferred in elastic scattering scales with\break $(A+1)^{-1}$, hydrogenous materials are the most
efficient neutron moderators. In sampling calorimeters with hydrogenous active material, the recoil protons may contribute to the calorimeter signals. Acosta \etal ~have  demonstrated that the average time between subsequent 
elastic scattering processes in that case is approximately constant, and since the energy fraction transferred to the protons is, on average, constant as well (50\%), this contribution manifests itself as an exponential tail to the time structure of the signals \cite{Aco91a}. This is the signature we have looked for in the present study.

In Reference \cite{Aco91a}, the time constant of the exponential tail was measured to be 9.9 ns. 
The time constant in the detector used for our studies may be estimated as follows. Hydrogen atoms
were contained in the plastic fibers that yielded the signals in this detector. These fibers constituted
17.4\% of the total detector volume, which further contained copper (69.3\%), silicon (4.6\%) and air (8.7\%).
Assuming that plastic contains equal numbers of hydrogen and carbon atoms, and has a density 
of 1 g/cm$^3$, we calculated a density of $8\cdot 10^{21}$ hydrogen atoms per cm$^3$.

The cross section for elastic neutron-proton scattering in the relevant energy range increases from 2.2 b at 3 MeV to 12 b at 0.1 eV \cite{sigmanp}. Therefore, the average distance traveled by neutrons between subsequent $np$ scattering processes varied from 56 cm at 3 MeV to 10 cm at 0.1 MeV, while the velocity of the neutrons decreased from 0.08$c$ at 3 MeV to 0.015$c$ at 0.1 MeV. As a result, the {\sl time} between subsequent $np$ scatters was the same for these two energies: 23 ns. This is of course the reason for the exponentially decreasing contribution of the neutrons to the signal. In each $np$ scatter, the neutrons lose, on average, the same fraction of their energy: 50\%. If no other energy loss mechanisms than elastic $np$ scattering would play a role, the kinetic energy of the neutrons would thus decrease by a factor $e$ in 33 ns. Other, competing energy loss processes would speed up this decrease.
Such processes include elastic scattering off C, Si and Cu nuclei, as well as inelastic $(n,n'\gamma)$
scattering, mainly off $^{63,65}$Cu nuclei (lowest excited states $\sim 0.7$ MeV). We estimate these processes to decrease the time constant of the neutron tail to $\sim 25$ ns.

As indicated above, the mean free path of the evaporation neutrons is quite large. Neutrons produced 
at 3 MeV travel on average half a meter before undergoing their first scatter off a free proton. This means 
that the radial profile of the neutron contribution to the signals is much broader than that of the other shower particles, whose profile is governed by the Mol\`iere radius (for the em shower component)  and the nuclear interaction length. Therefore, the relative contribution of neutrons to the scintillation signals should increase with the radial distance to the shower axis.
Since the recoil protons are non-relativistic, they do not generate \v{C}erenkov light.  In summary, the signature we are looking for thus has the following characteristics:
\begin{itemize}
\item An exponential tail in the time structure of the signals, with a time constant of $\sim 25$ ns
\item The relative importance of this tail, \ie of its contribution to the total calorimeter signal, should increase with the distance to the shower axis
\item The tail should be absent in the time structure of calorimeter signals based on \v{C}erenkov light
\end{itemize}
In the following, we describe our search for an experimental signature of this type.

\section{Experimental details}

\subsection{The detector}

\begin{figure}[htb]
\epsfysize=7cm
\centerline{\epsffile{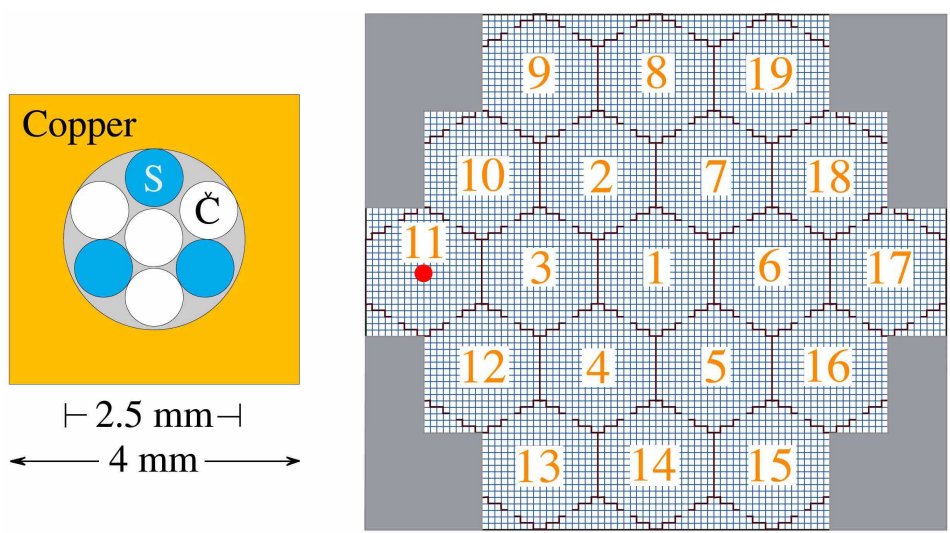}}
\caption{\small
The basic building block of the DREAM calorimeter is a 4 $\times$ 4 mm$^2$ extruded hollow copper rod of 2 meters length, with a 2.5
mm diameter central hole. Seven optical fibers\break (4 undoped and 3 scintillating fibers) with a diameter of 0.8 mm each are
inserted in this hole, as shown in the left diagram. The right diagram shows a cross section of the calorimeter, which consists of 19 hexagonal towers. The impact point of the beam (center of tower \#11) is indicated as well.}
\label{layout}
\end{figure}

The detector used for our studies was the DREAM calorimeter, which has been described in considerable detail elsewhere \cite{DREAMem,DREAMhad,DREAMmu}.
The basic element of this detector (see Figure \ref{layout}) is an extruded copper rod, 2 meters long
and 4$\times$ 4 mm$^2$ in cross section. This rod is hollow, and the central cylinder has a diameter of 2.5 mm. 
Seven optical fibers are inserted in this hole. Three of these are plastic scintillating fibers, the other four fibers are undoped and are intended for detecting \v{C}erenkov light. 

The DREAM detector consisted of 5580 such rods, 5130 of these were equipped with fibers. 
The empty rods were used as fillers, on the periphery of the detector. The instrumented volume thus had a
length of 2.0 m, an effective radius of 16.2 cm, and a mass of 1030 kg. The calorimeter's radiation
length ($X_0$) was 20.1 mm, its Moli\'ere radius ($\rho_M$) 20.4 mm and its nuclear interaction length ($\lambda_{\rm int}$) 200 mm.

The fibers were grouped to form 19 readout towers. Each tower consisted of 270 rods and had an approximately hexagonal
shape (80~mm apex to apex). The effective radius of each tower was 37.1 mm ($1.82 \rho_M$).
A central tower (\#1) was surrounded by 2 hexagonal rings, the Inner Ring (6 towers, numbered 2 - 7) and the Outer Ring (12 towers, numbered 8 - 19). 
The towers were
not segmented in the longitudinal direction.

The fibers leaving the rear of this
structure were separated into bunches: one bunch of scintillating fibers and one bunch of \v{C}erenkov fibers for each tower, 
38 bunches in total. In this way, the readout structure was established (see Figure \ref{layout}). Each bunch was coupled
through a 2 mm air gap to a photomultiplier tube (PMT)\footnote{Hamamatsu R-580, a 10-stage,
1.5" PMT with a nominal gain of $3.7\cdot 10^5$ at 
1250 V.}. Much more information about this calorimeter is provided in References \cite{DREAMem,DREAMhad,DREAMmu}.

\subsection{The beam line}

The measurements described in this paper were performed in the H4 beam line of the Super Proton Synchrotron
at CERN. The DREAM detector
was mounted on a platform that could move vertically and sideways with respect to the beam.
For the measurements described here, we only used one detector position, namely where the
beam entered the detector parallel to its axis (the ``$0^\circ$" orientation), in the center of Tower \#11.

Two small scintillation counters provided the signals that were used to trigger the data acquisition system.
These Trigger Counters were 2.5 mm thick, and the area of overlap was 6$\times$6 cm$^2$. A coincidence between the logic
signals from these counters provided the trigger.

\subsection{Data acquisition}

Measurement of the time structure of the calorimeter signals formed a crucial part of the tests
described here. In order to limit distortion of this structure as much as possible, we used 15 mm thick air-core cables to transport 4 selected calorimeter signals to the counting room. Such cables were also used for
the trigger signals, and these were routed such as to
minimize delays in the DAQ system\footnote{We measured the signal speed to be 0.78$c$ in these cables.}.

The  other calorimeter signals  were transported through RG-58 cables with (for timing purposes) appropriate lengths to the counting room.
The  signals used for the neutron measurements were split (passively) into 3 equal parts in the counting room. One part was sent to a charge ADC, the other 2 signals were used for analysis of the time structure by means of a FADC. The latter unit measured the amplitude of the signals at a rate of 200 MHz.
During a time interval of 80 ns, 16 measurements of the amplitude were thus obtained. The 2 signals from the splitter box were measured separately in 2 different channels of the FADC module\footnote{Dr. Struck SIS3320,  http://www.struck.de/sis3320.htm}. The second signal was delayed by 2.5 ns with respect to the first one. 
By using 2 channels of the FADC module for each calorimeter signal, the time structure of the signals was thus effectively measured with a resolution of 2.5 ns (400 MHz).

The charge measurements were performed with 12-bit LeCroy 1182 ADCs.
These had a sensitivity of 50 fC/count and a conversion time of 16 $\mu$s.
The ADC gate width was 100 ns, and the calorimeter signals arrived $\sim 20$ ns after the start of the gate.

The data acquisition system used VME electronics.
A single VME crate hosted all the needed readout and control boards.
The trigger logic was implemented through NIM modules and the signals were sent 
to a VME I/O register, which also collected the spill and the global 
busy information. The VME crate was linked to a Linux based computer 
through an SBS 620\footnote{http://www.gefanucembedded.com/products/457} 
optical VME-PCI interface that allowed memory 
mapping of the VME resources via an open source driver\footnote{http://www.awa.tohoku.ac.jp/$\sim$sanshiro/kinoko-e/vmedrv/}. The computer was equipped with a 2 GHz Pentium-4 
CPU, 1 GB of RAM, and was running a CERN SLC 4.3 operating system\footnote{http://linux.web.cern.ch/linux/scientific4/}.

The data acquisition was based on a single-event polling mechanism and 
performed by a pair of independent programs that communicated
through a first-in-first-out buffer, built on top of a 32 MB shared 
memory. Only exclusive accesses were allowed and concurrent requests were 
synchronised with semaphores. The chosen scheme 
optimized the CPU utilization and increased the data taking efficiency by 
exploiting the bunch structure of the SPS, where beam particles were provided to
our experiment during a spill of 4.8 s, out of a total cycle time of 16.8 s.
During the spill, the readout program collected data from the VME modules and 
stored them into the shared memory, with small access times. During the remainder of the SPS cycle, a 
recorder program dumped the events to the disk. Moreover, the buffer
presence allowed low-priority monitoring programs to run (off spill) in 
spy mode. With this scheme, we were able to reach a data acquisition rate 
as high as 2 kHz, limited by the FADC readout time. 
The typical event size was $\sim 1$ kB.  
All calorimeter signals and the signals from the auxiliary detectors were monitored on-line.

\subsection{Calibration of the detectors}

Using the high voltage, the gain in all PMTs was set to generate 
$\sim 1$~pC/GeV.
The 38 PMTs reading out the 19 towers were calibrated with 50 GeV
electrons. The showers generated by these particles were not completely contained in a single calorimeter tower. The
(average) containment was found from EGS4 Monte Carlo simulations. 
When the electrons entered a tower in its geometrical center, on average 
$92.5\%$ of the scintillation light and $93.6\%$ of the \v{C}erenkov light was generated in that tower \cite{DREAMem}.
The remaining fraction of the light was shared by the surrounding towers.
The signals observed in the exposed tower thus corresponded to an energy deposit of 46.3 GeV in the case of the 
scintillating fibers and of 46.8 GeV for the \v{C}erenkov fibers. 
This, together with the precisely measured values of the average signals from the exposed tower, formed the basis for determining 
the calibration constants, \ie the relationship between the measured number of ADC counts and the corresponding energy deposit.   

\subsection{Experimental data}

The experiments were carried out with a beam of 100 GeV $\pi^+$ which was steered into the center of Tower \#11.
In all measurements, the scintillation and \v{C}erenkov signals from this (on-axis) tower were sent through the
air-core cables for time structure analysis, using $2\times 2 = 4$ channels of the FADC unit. The other 4 FADC channels were used to measure the time structure of the signals from an off-axis tower. In separate runs, we used
the signals from Tower \#3 (located at an average radial distance of 72 mm from the beam axis), Tower \#1 (radial distance 144 mm) and Tower \#6 (radial distance 216 mm) for that purpose. In each run, 100 000 events were
collected.

\section{Experimental results}

\begin{figure}[htb]
\epsfysize=12cm
\centerline{\epsffile{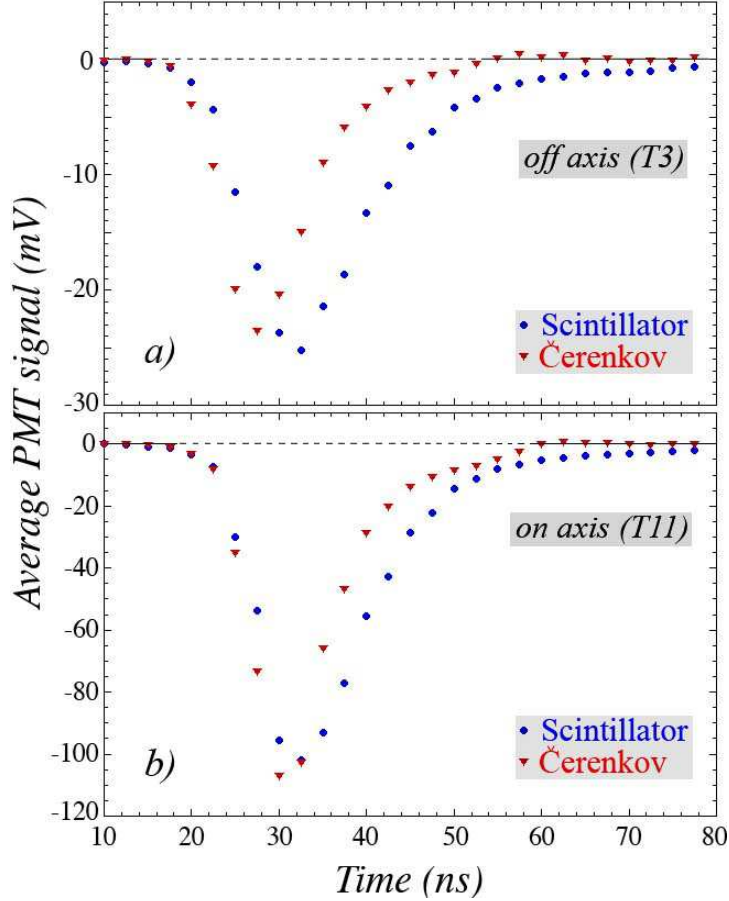}}
\caption{\small
Average time structure of the scintillator and \v{C}erenkov signals measured for 100 GeV $\pi^+$ showers in a DREAM tower (\#11) located on the shower axis ($b$) and in a tower (\#3) located at a distance of 72 mm off-axis ($a$). }
\label{Q-S11}
\end{figure}

The FADC measurements provided considerable detail on the time structure of the signals generated by the 
calorimeter. This is illustrated in Figure \ref{Q-S11}, which shows the average time structure of the scintillator
and \v{C}erenkov signals measured in the neighboring towers \#11 (on-axis) and \#3 (off-axis), with a sampling frequency of 400 MHz (2.5 ns time samples).
Two general features should be pointed out:
\begin{itemize}
\item The starting point of the pulse is not the same for  all 4 signals, as a result of differences in cable lengths
and PMT transit times. In particular, the signals from Tower \#3 started about 2.5 ns earlier than those from Tower 
\#11.
\item The input impedances of the channels that recorded the signals from the off-axis tower were four times smaller than those of the channels that handled the signals from the on-axis tower. Therefore, the vertical scale
of the bottom plot should be multiplied by a factor of 4 in order to be compatible with the top one.
\end{itemize}
The figure also shows that the \v{C}erenkov signals are considerably faster than the scintillation ones. This should of course be expected, since the scintillation process is characterized by one or several time constants, while \v{C}erenkov light emission is prompt. However, for our purpose it  is more interesting to compare the corresponding signals from the two different calorimeter towers with each other. This is because the relative contribution of neutrons to the signals increases with the distance to the shower axis.
\begin{figure}[htb]
\epsfysize=7.5cm
\centerline{\epsffile{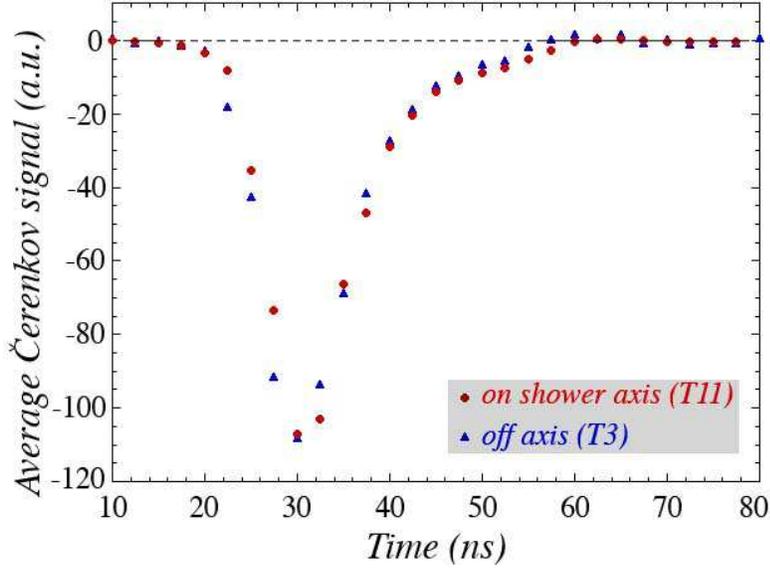}}
\caption{\small
Comparison of the average time structure of the \v{C}erenkov signals measured for 100 GeV $\pi^+$ showers in a DREAM tower (\#11) located on the shower axis and in a tower (\#3) located at a distance of 72 mm off-axis. The signals have been normalized such that the integral over the pulse shape is the same in both cases.}
\label{Qcomparison}
\end{figure}

Figure \ref{Qcomparison} shows the time structure of the \v{C}erenkov signals from Towers \#11 and \#3 in one
plot. The time axis of the Tower \#3 signal has been shifted by 2.5 ns to make the starting points of both signals
the same. In order to be able to compare the pulse shapes, we have normalized both signals on the basis of their
integrated pulse shape. The result shows no significant differences between the time structures of the \v{C}erenkov signals from these two towers. Since the low-energy neutrons do not contribute to the \v{C}erenkov signals from
this calorimeter other than through capture $\gamma$s (which fall outside the 80 ns time scale considered here),
we did not expect to see a difference.
\begin{figure}[htb]
\epsfysize=8.5cm
\centerline{\epsffile{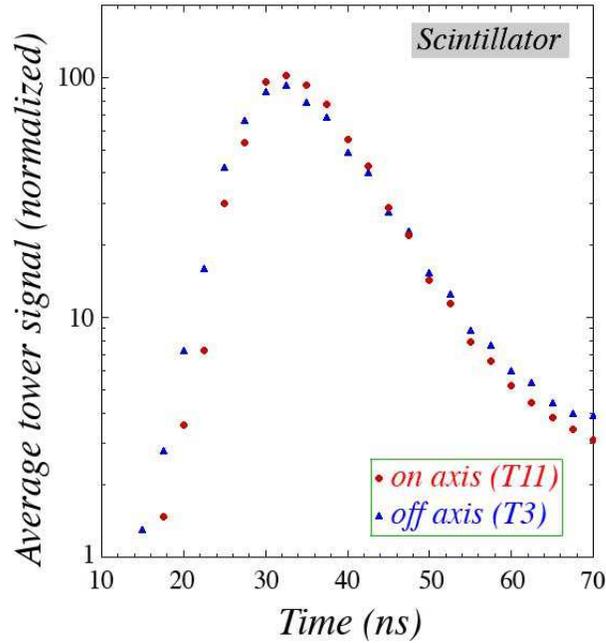}}
\caption{\small
Comparison of the average time structure of the scintillator signals measured for 100 GeV $\pi^+$ showers in a DREAM tower (\#11) located on the shower axis and in a tower (\#3) located at a distance of 72 mm off-axis). The measured signals have been inverted.}
\label{S3-S11log}
\end{figure}

The situation is quite different for the scintillation signals. Figure \ref{S3-S11log} shows the average time structure of the scintillation signals from Towers \#11 and \#3 in a logarithmic display. As in Figure \ref{Qcomparison}, the integrated pulse shapes have been equalized in order to facilitate a comparison between these time structures.
The figure shows that the trailing edge of the signal from Tower \#11 is, on average, clearly steeper than that of the Tower \#3 pulse. Since an eventual contribution of neutrons to the scintillator signals
is expected to increase with the distance to the shower axis, this is precisely the effect one would expect to see.
\begin{figure}[htb]
\epsfysize=8cm
\centerline{\epsffile{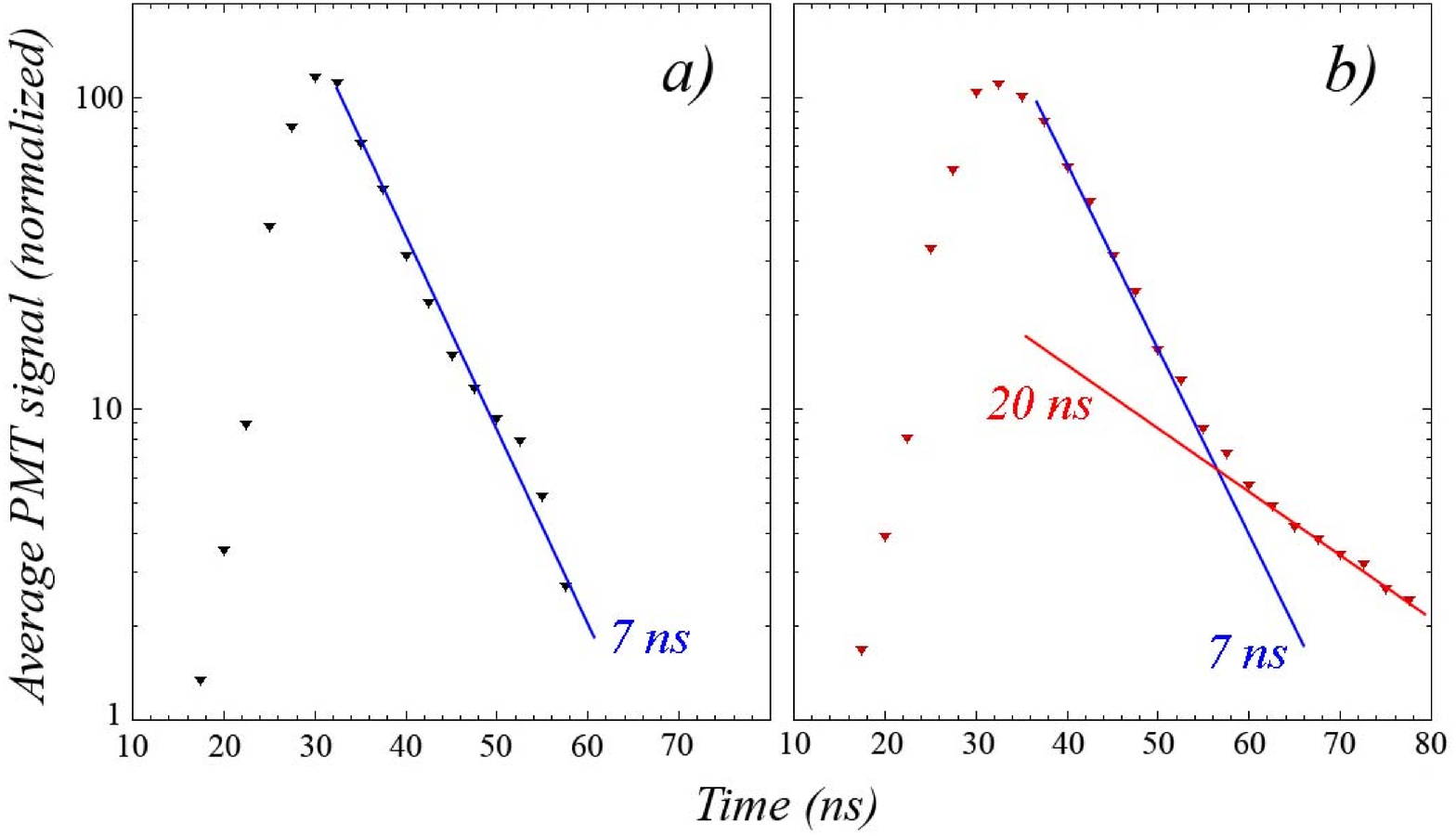}}
\caption{\small
The average time structure of the \v{C}erenkov ($a$) and  scintillator ($b$) signals measured for 100
GeV $\pi^+$ showers in DREAM tower \#11, located on the shower axis. The measured signals have 
been inverted. The lines represent exponential fits to (parts of) the trailing edge of the signal shapes.}
\label{SQ11}
\end{figure}

Figure \ref{SQ11} shows a comparison between the average \v{C}erenkov and scintillator signals from Tower \#11. The signals have been inverted and are logarithmically displayed, so that differences in 
the signal shapes are emphasized. The trailing edge of the \v{C}erenkov pulse shape is well described by a single exponential function, with a time constant of $\sim 7$ ns. Such a function also describes the
initial portion of the trailing edge of the scintillator pulse shape. However, the latter shape exhibits clearly
an additional, slower component. The curve drawn in Figure \ref{SQ11}b corresponds to a time constant of 20 ns for this slow component.

We fitted the trailing edge of the average scintillator signal distributions observed in Towers \#11, 3, 1 and 6 to an expression of the following type:
\begin{equation}
N(t) = N_1 e^{-t/\tau_1} ~+~N_2 e^{-t/\tau_2}
\label{int}
\end{equation}
where the decay time constants $\tau_1$ and $\tau_2$ were kept at fixed values, and the signal values $N_1$ and $N_2$ were optimized in the fit. The ratio $N_2/N_1$ is a measure for the relative contribution 
of neutrons to the scintillator signals. The precise value of this contribution was found by calculating
\begin{equation}
f_n ~=~{{\int_{t_0}^\infty{N_2 e^{-t/\tau_2} dt}}\over {\int_{t_0}^\infty{(N_1e^{-t/\tau_1}~+~N_2 e^{-t/\tau_2}) dt}}}
\end{equation}
where $t_0$ is chosen such that 
\begin{equation}
\int_{t_0}^\infty{(N_1e^{-t/\tau_1}~+~N_2 e^{-t/\tau_2}) dt} = \Sigma_i S_i
\end{equation}
the experimentally measured integrated pulse shape, shown in Figure \ref{SQ11}b. 
The value of $\tau_1$ was kept constant at 7 ns throughout these studies, whereas for $\tau_2$ values
of 20, 25 and 33 ns were used. Because of the limited time interval over which the time structure was measured, and the relatively small contribution of the slow component to the total signals, it was not
possible to measure $\tau_2$ with very high precision, although the value of 25 ns estimated in our analysis described in Section 2.2 is most definitely not far off the mark.
\begin{figure}[htb]
\epsfysize=8.5cm
\centerline{\epsffile{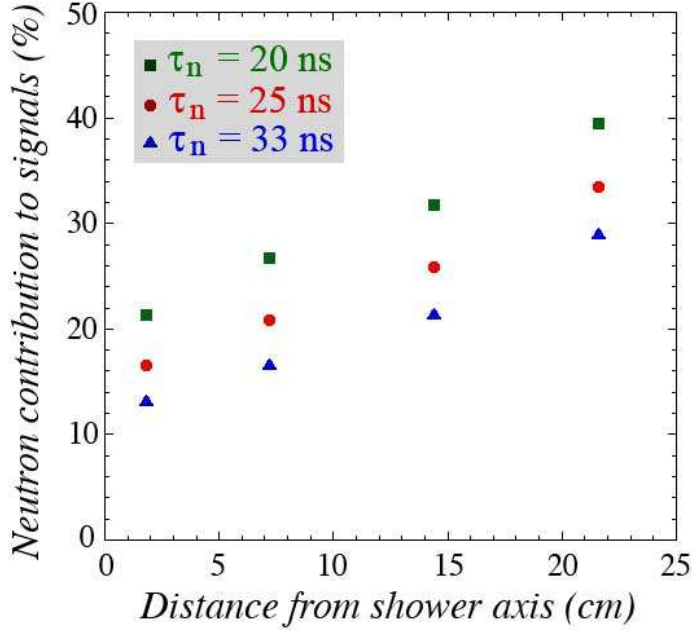}}
\caption{\small
The average percentage of the scintillation signals from 100 GeV $\pi^+$ showers attributed to elastic neutron-proton scattering in the plastic fibers, for four different DREAM towers, located at  different, increasing distances from the shower axis. Results are given for 3 different choices of the time constant assumed for this neutron contribution. See text for details. }
\label{neutrons}
\end{figure}
Yet, the results of this analysis show clearly that
the relative contribution of neutrons to the total scintillator signals ($f_n$) increases with the distance from the shower axis. This is illustrated in Figure \ref{neutrons}, where $f_n$ is shown for Towers \#11, 3, 1 and 6.
For each of the 3 mentioned values of the decay time constant $\tau_2$, the average neutron contribution
approximately doubles, from $\sim 15\%$ to $\sim 30\%$, when going out from the shower axis (the center of Tower \#11)  to Tower \#6, whose center is located at 21.6 cm from the shower axis.

\section{Conclusions}

We have analyzed the time structure of hadronic signals from the DREAM calorimeter, with a sampling frequency
of 400 MHz (2.5 ns time bins), for towers located at different distances from the shower axis.
We have found a clear indication for the contribution of evaporation neutrons to the scintillation signals from this detector. These neutrons contribute through
elastic scattering off protons in the plastic scintillating fibers which provide the $dE/dx$ information in this calorimeter. Their contribution is characterized by an exponential tail in the pulse shape, with a time
constant of $\sim 25$ ns. The contribution of neutrons to the signals increases with the distance from the shower axis, and 
represents up to $\sim 30\%$ of the total signal from off-axis towers.
As expected, the neutrons do not contribute at all to the DREAM \v{C}erenkov signals.

\section*{Acknowledgments}

We thank CERN for making particle beams of excellent quality available.
This study was carried out with financial support of the United States
Department of Energy, under contract DE-FG02-95ER40938.

\bibliographystyle{unsrt}

\end{document}